\documentclass[12pt]{article}

\usepackage{amsmath}
\usepackage[dvipsnames]{xcolor}
\usepackage{graphicx,psfrag,epsf}
\usepackage{enumerate}
\usepackage{natbib}
\usepackage{algorithm}
\usepackage{algorithmicx}
\usepackage{algpseudocode}
\usepackage{url} 
\usepackage{mathtools,amssymb,amsthm}
\usepackage{soul} 
\usepackage{float} 
\usepackage[colorlinks=true,linkcolor=blue,urlcolor=NavyBlue,citecolor=RubineRed]{hyperref}
\usepackage{booktabs}
\usepackage{multirow}
\usepackage{kotex}
\usepackage{color, colortbl}
\usepackage{caption}
\usepackage{floatpag}

\usepackage{todonotes} 
\usepackage{comment} 

\newif\ifshow 
\showfalse 

\ifshow
  \includecomment{wrap}
\else
  \excludecomment{wrap} 
\fi

\newcommand{\beq}{\begin{eqnarray}}
\newcommand{\eeq}{\end{eqnarray}}

\newcommand{\bd}{\mathbf{ d}}

\newcommand{\bm}{\mathbf{ m}}

\newcommand{\bM}{\mathbf{ M}}

\newcommand{\bX}{\mathbf{ X}}

\newcommand{\balpha}{\boldsymbol{\alpha}}
\newcommand{\bbeta}{\boldsymbol{\beta}}
\newcommand{\bgamma}{\boldsymbol{\gamma}}
\newcommand{\bdelta}{\boldsymbol{\delta}}

\newcommand{\blambda}{\boldsymbol{\lambda}}

\newcommand{\btau}{\boldsymbol{\tau}}

\newcommand{\bomega}{\boldsymbol{\omega}}

\newcommand{\bzero}{{\mathbf 0}}
\newcommand{\bone}{{\mathbf 1}}
\def\red{\color{red}}
\def\blue{\color{blue}}

\def\spacingset#1{\renewcommand{\baselinestretch}%
{#1}\small\normalsize} \spacingset{1}

\theoremstyle{plain}

\theoremstyle{definition}

\newcommand{\titlefont}{\fontsize{17}{22}\selectfont\bfseries}

\begin{document}

\title{\titlefont Bayesian Variable Selection for High-Dimensional Mediation Analysis: Application to Metabolomics Data in Epidemiological Studies}

\author{Youngho Bae$^{1}$, Chanmin Kim$^{1}$, Fenglei Wang$^{2}$, Qi Sun$^{2,3,4}$, Kyu Ha Lee$^{2,3,5,\ast}$\\ \\
    \textit{\small $^{1}$Department of Statistics, Sungkyunkwan University, Seoul, South Korea}\\
    \textit{\small $^{2}$Department of Nutrition, Harvard T.H. Chan School of Public Health, Boston, MA, U.S.A.}\\
    \textit{\small $^{3}$Department of Epidemiology, Harvard T.H. Chan School of Public Health, Boston, MA, U.S.A.}\\
    \textit{\small $^{4}$Channing Division of Network Medicine, Brigham and Women's Hospital, Boston, MA, U.S.A.}\\
    \textit{\small $^{5}$Department of Biostatistics, Harvard T.H. Chan School of Public Health, Boston, MA, U.S.A.}\\
    {\small $^\ast$klee@hsph.harvard.edu}}
    
    \date{}
    
\maketitle

\begin{abstract}
\noindent In epidemiological research, causal models incorporating potential mediators along a pathway are crucial for understanding how exposures influence health outcomes. This work is motivated by integrated epidemiological and blood biomarker studies, investigating the relationship between long-term adherence to a Mediterranean diet and cardiometabolic health, with plasma metabolomes as potential mediators. Analyzing causal mediation in such high-dimensional omics data presents substantial challenges, including complex dependencies among mediators and the need for advanced regularization or Bayesian techniques to ensure stable and interpretable estimation and selection of indirect effects. To this end, we propose a novel Bayesian framework for identifying active pathways and estimating indirect effects in the presence of high-dimensional multivariate mediators. Our approach adopts a multivariate stochastic search variable selection method, tailored for such complex mediation scenarios. Central to our method is the introduction of a set of priors for the selection indicators in the mediator and outcome models: a Markov random field prior and sequential subsetting Bernoulli priors. The first prior's Markov property leverages the inherent correlations among mediators, thereby increasing power to detect mediated effects. The sequential subsetting aspect of the second prior encourages the simultaneous selection of relevant mediators and their corresponding indirect effects from the two model parts, providing a more coherent and efficient variable selection framework, specific to mediation analysis. Comprehensive simulation studies demonstrate that the proposed method provides superior power in detecting active mediating pathways. We further illustrate the practical utility of the method through its application to metabolome data from two sub-studies within the Health Professionals Follow-up Study and Nurses’ Health Study II, highlighting its effectiveness in real data setting.
\end{abstract}

\noindent%


\spacingset{1.45} 


\section{Introduction} \label{sec_intro}

Mediation analysis is a powerful tool in epidemiological research, allowing for a deeper understanding of the biological mechanisms through which exposures influence health outcomes. In studies that integrate blood biomarkers, the use of high-dimensional omics measures as mediators offers a unique opportunity to identify the pathways between exposures and health conditions. For instance, in our motivating epidemiological studies exploring the association between long-term adherence to a Mediterranean diet and cardiometabolic health, metabolomic profiles can serve as key mediators by capturing the small-molecule metabolites that reflect biochemical processes triggered by diet. 

This work is motivated by two substudies within the Health Professionals Follow-up Study (HPFS) and Nurses' Health Study II (NHSII). The HPFS and the NHSII are ongoing prospective cohort studies involving 51,529 U.S. male health professionals, initiated in 1986 \citep{rimm1991prospective}, and 116,429 female registered nurses, initiated in 1989 \citep{bao2016origin}, respectively. Detailed participant information on diet, lifestyle, medication use, and health outcomes was collected at baseline and updated biennially. The substudies include 307 men and 233 women, all free from coronary heart disease, stroke, cancer, or major neurological diseases, and involve the collection of blood samples from participants \citep{huang2019mind, wang2021gut}. Metabolomics data derived from blood biomarkers have been extensively analyzed in impactful studies, including investigations into the association between plasma metabolite profiles linked to plant-based diets and type 2 diabetes \citep{wang2022plasma}, and the pivotal role of these profiles in mediating the relationship between haem iron intake and type 2 diabetes risk \citep{wang2024integration}. Our analysis in this work focuses on (i) identifying a subset of metabolites that mediate the relationship between adherence to a Mediterranean diet and cardiometabolic health outcomes, and (ii) estimating and making inferences about the indirect effect sizes associated with these metabolites.

\begin{figure}[ht]
\centering
\includegraphics[width=9cm]{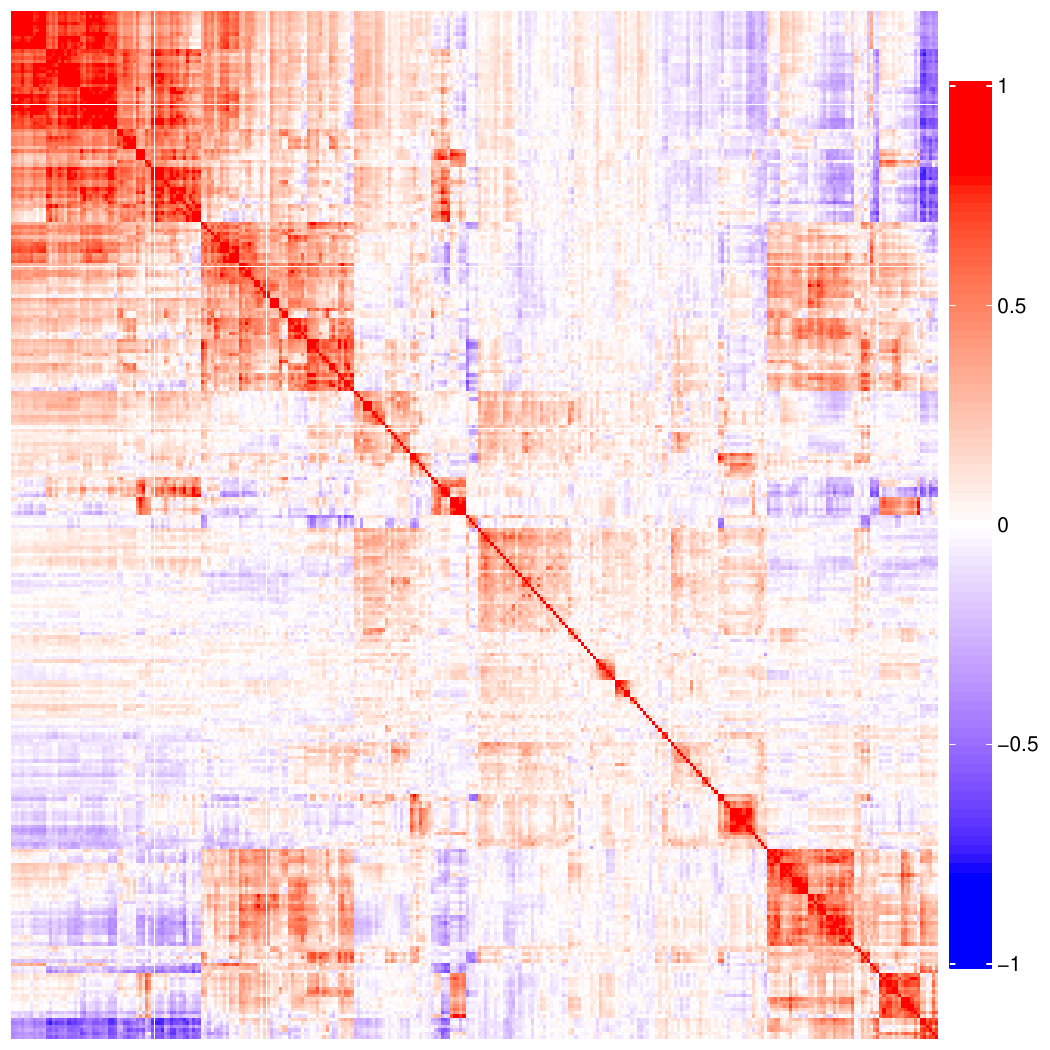}
\caption{Empirical correlations among the measures of {\blue 298} metabolites in the substudies within the Health Professionals Follow-up Study (HPFS) and Nurses' Health Study II (NHSII).} \label{fig:corr}
\end{figure}

Broadly, published methodologies for mediation analysis with high-dimensional mediators can be categorized into three main approaches: methods employing dimension reduction techniques, such as principal component analysis and its variants \citep{huang2016hypothesis, chen2018high, zhao2020sparse}, as well as latent mediator models \citep{derkach2019high}; penalization-based methods that apply regularization techniques to reduce the set of potential mediators before mediator-outcome analysis \citep{zhang2016estimating, gao2019testing, perera2022hima2, zhang2022high, zhao2022pathway}; and joint modeling frameworks, which estimate exposure-mediator-outcome pathways simultaneously, addressing uncertainties across all stages of estimation \citep{aung2020application, song2020bayesianOmics, song2021bayesian, song2021bayesianSparse}. Each strategy has certain limitations. For instance, dimension reduction techniques condense information from multiple mediators but restrict the ability to interpret mediation effects at the individual mediator level. Similarly, penalization-based methods, which follow a two-stage approach, do not adequately account for uncertainty estimated from the selection of mediators, potentially biasing downstream mediator-outcome inference. In contrast, joint modeling frameworks overcome these limitations by unifying pathway estimation; however, the existing methods assume independent mediators or employ simple latent neighboring structures with limited connectivity between mediators \citep{song2021bayesian}, likely due to the challenges of computational complexity. Moreover, a common drawback across these approaches is their tendency to overlook correlations among mediators (Figure \ref{fig:corr}). As demonstrated in the multivariate analysis literature \citep{breiman1997predicting, saccenti2014reflections, lee2017multivariate, lee2020bayesian}, ignoring these dependencies can lead to inflated false positive rates, reduced statistical power, and decreased predictability. To our knowledge, no existing statistical methods for joint mediation analysis leverage mediator correlations through specialized priors or penalty terms to enhance the detection of active mediating pathways.

Toward overcoming these limitations, we propose a novel Bayesian mediation analysis method that identifies active exposure–mediator–outcome pathways and estimates the corresponding indirect effects within a joint modeling framework, while accommodating correlations among high-dimensional mediators. The foundation of our approach lies in the introduction of a Markov random field (MRF) and a conditional Bernoulli prior for latent binary selection indicators in the mediator and outcome models, respectively. The application of these priors in mediation analysis offers advantages in two key aspects. First, it leverages the covariance structure among mediators by specifying edge potentials within the MRFs. This structure is adaptively updated at each step of the Markov chain Monte Carlo (MCMC) algorithm, enabling variable selection to account for both the mean model and variance-covariance structure of the mediators. Second, the sequential subsetting nature of the conditional Bernoulli prior promotes simultaneous selection of relevant mediators and their indirect effects across both mediator and outcome models, leading to a more coherent and efficient variable selection procedure for mediation analysis.

The remainder of this article is organized as follows: Section \ref{sec:mediationa analysis} outlines the high-dimensional mediation analysis framework, including observed data model and identifying assumptions; Section \ref{sec:bvs} details the proposed Bayesian framework, including the specification of prior distributions and an overview of an efficient computational algorithm for sampling from the joint posterior distribution; Section \ref{sec:simulation} presents comprehensive simulation studies assessing the performance of the proposed framework; Section \ref{sec:application} provides an detailed analysis of the motivating NHSII/HPFS substudies; and Section \ref{sec:discussion} concludes the article with a discussion. Supplementary details, including computational algorithms and additional results, are provided in an online materials document as appropriate.

\begin{wrap}

{\red When analyzing high-dimensional mediators, dimension reduction methods are often employed to transform these mediators into a smaller set of independent variables. \cite{huang2016hypothesis} used principal component analysis (PCA) to reduce correlated mediators into a lower-dimensional set of uncorrelated variables, allowing for testing individual pathway effects. However, interpreting the reduced dimensions can be challenging. To address this, \cite{zhao2020sparse} utilized sparse PCA \citep{zou2006sparse}, which applies regularization to sparsify the loading vector, thereby improving the interpretability of the transformed mediators. Despite these advancement, obtaining a direct interpretation of mediation effects from the original mediators remains difficult. \cite{chen2018high} conducted their analysis by reducing the dimensionality of the mediators through a linear combination of the original mediators that are orthogonal to each other. \cite{derkach2019high} introduced latent variables associated with the mediators to estimate the mediation effect. However, these methods also share the limitation of making it difficult to interpret the results in terms of the original mediators.

Several approaches have been proposed to reduce the set of potential mediators for analysis using penalized methods. \cite{zhang2016estimating} and \cite{perera2022hima2} suggested selecting a subset of mediation pathways through sure independence screening (SIS, \cite{fan2008sure}), followed by further analysis. Under the assumption that mediators are uncorrelated, \cite{zhang2022high} introduced mediation pathway selection using scaled adaptive lasso, while \cite{zhao2022pathway} developed pathway lasso, a novel penalty aimed at minimizing the number of mediators required for inference. In contrast, \cite{gao2019testing} proposed a simultaneous inference procedure that accounts for correlations among selected mediators. However, a key limitation of these methods is their inability to incorporate the uncertainty from each step into the final results.

Bayesian approaches have been proposed to account for uncertainties in all stages of estimation. These methods typically involve specifying mediator models using a multivariate normal distribution and employing Bayesian variable selection techniques to identify significant exposure-mediator-outcome pathways \citep{aung2020application,song2021bayesianSparse}. However, a key limitation is the assumption that mediators are uncorrelated. \cite{song2020bayesianOmics} introduced a method for selecting mediators using a mixture of two normals prior, without imposing restrictions on the correlation structure among multivariate mediators. Nonetheless, the method was developed under the assumption of independence among mediators. In contrast, \cite{song2021bayesian} estimated individual mediation effects based on latent neighboring relationships among mediators, though the approach did not explicitly specify the correlation structure. As a result, these methods generally overlook correlations among mediators (e.g., metabolites) (Figure \ref{fig:corr}), which compromises critical information and, as highlighted in the multivariate analysis literature, leads to inflated type I error rates and diminished statistical power \citep{breiman1997predicting, saccenti2014reflections, lee2017multivariate, lee2020bayesian}.

In this manuscript, we propose.... 
}

{\blue A paragraph discussing existing possible mediation analysis frameworks and corresponding references, which are limited in that they i) ignore correlations among mediators, and/or ii) are unable to provide interpretable indirect effects at individual mediator level, and/or iii) are not applicable for high-dimensional mediators (no-VS for e.g.) and/or iv) come with other limitations.}

{\blue [Insert the following sentence after i) above:]} These strategies typically ignore the correlations among mediators (i.e. metabolites) (Figure \ref{fig:corr}), thus sacrificing information and, as established in multivariate analysis literature, leading to highly inflated type I error rates and reduced statistical power \citep{breiman1997predicting, saccenti2014reflections, lee2017multivariate, lee2020bayesian}.

\begin{itemize}
    \item Begin with a concise introduction to the topic we will be discussing (i.e., bvs+high-dimensional mediation). 
    \item Review the current literature on high-dimensional mediation analysis.
    \begin{itemize}
        \item Bayesian hierarchical models for high‐dimensional mediation analysis with coordinated selection of correlated mediators \citep{song2021bayesian}
        \item Application of an analytical framework for multivariate mediation analysis of environmental data \citep{aung2020application}
        \item Bayesian sparse mediation analysis with targeted penalization of natural indirect effects \citep{song2021bayesianSparse}
        \item Bayesian shrinkage estimation of high dimensional causal mediation effects in omics studies \citep{song2020bayesianOmics}
        \item HIMA2: high-dimensional mediation analysis and its application in epigenome-wide DNA methylation data \citep{perera2022hima2}
        \begin{itemize}
            \item high-dimensional.. but univariate analysis… leaves correlated mediators as future  work.. “HIMA2 lose power for high correlation.”
        \end{itemize}
        \item Instrumental variable-based high-dimensional mediation analysis with unmeasured confounders for survival data in the observational epigenetic study\citep{chen2023instrumental}
        \begin{itemize}
            \item “A2. The mediators are independent of each other.” survival outcomes
        \end{itemize}
        item Mediation analysis for survival data with high-dimensional mediators \citep{zhang2021mediation}
        \begin{itemize}
            \item survival outcomes. looks as if they handle the correlation structure of Sigma\_e but--``Step 1:(Mediators screening). Motivated by the sure independence screening (SIS) (Fan and Lv, 2008; Fan et al., 2010), we consider a series of marginal models"
        \end{itemize}
        \item High dimensional mediation analysis with latent variables \citep{derkach2019high,chen2018high}
        \begin{itemize}
            \item Factor-analytic model: more general than our model in that it allows multiple factors. However, indirect effects are only interpretable at the level of latent variables (not at the actual mediators level)
        \end{itemize}
        \item Dimension reduction\citep{huang2016hypothesis,zhao2020sparse,zhou2020estimation}: Using dimension reduction methods to transform high-dimensional parameters into a small number of independent new parameters. The two models are then fitted using the newly reduced parameters to directly estimate the indirect effects. (indirect effects through individual parameters cannot be estimated, and only the overall indirect effect can be measured. In other words, it is not possible to identify the "active" parameters).
        \item Penalized method \citep{zhang2016estimating,gao2019testing,zhang2022high,zhao2022pathway,song2020bayesianOmics}: Assuming no interaction between parameters, the two models are fitted to estimate the indirect effects for each parameter. The total indirect effect is then obtained by summing these individual effects. During the fitting process, a penalized method is applied to select significant parameters, and only the significant indirect effects are calculated.
    \end{itemize}

    \item Review the current literature on Bayesian variable selection (BVS).
    \item Review the current literature on the application of BVS to high-dimensional mediation, if available. Here, we need to talk about the setting (high-dimensional mediators) we are focusing on (maybe, along with other existing settings).
    \item Outline the proposed method -- what are the statistical innovations in the framework and our contributions?
    \begin{itemize}        
        \item high-dimensional mediators
        \item variable selection (any methods in Bayesian or frequentist?)
        \item multivariate distribution for mediator model
        \item Bayesian -- enables to incorporate dependence structure of multivariate mediators into variable selection/inference through a special prior
        \item R package software, illustrative examples on the GitHub
    \end{itemize}
\end{itemize}

\end{wrap}


\section{High-dimensional Mediation Analysis}\label{sec:mediationa analysis}
Let $A_i$ represent an exposure for $i = 1, \cdots, n$. Within the potential outcomes framework \citep{rubin1974estimating}, define $\boldsymbol{M}_i(a) = \{M_{1,i}(a), M_{2,i}(a), \cdots, M_{q,i}(a)\}$ as the potential values of the $q$-dimensional mediators for the $i$-th observation under exposure level $A_i = a$. Additionally, let $Y_i(a, \bm)$ represent the potential outcome that would be observed under exposure status $a$ and mediator vector $\bm$. Assuming the commonly applied Stable Unit Treatment Value Assumption (SUTVA) \citep{rubin1990formal,forastiere2016identification}, each potential mediator and outcome can be linked to the observed data as follows: $M_{j,i}(A_i) = M_{j,i}$ for $j=1, \cdots,  q$ and $Y_i(A_i, \boldsymbol{M}_i) = Y_i$. This implies that the potential mediators and outcome at the actual treatment level $A_i$ received by the $i$-th unit are equivalent to the observed values.

\begin{figure}[ht]
\centering
\includegraphics[width=10cm]{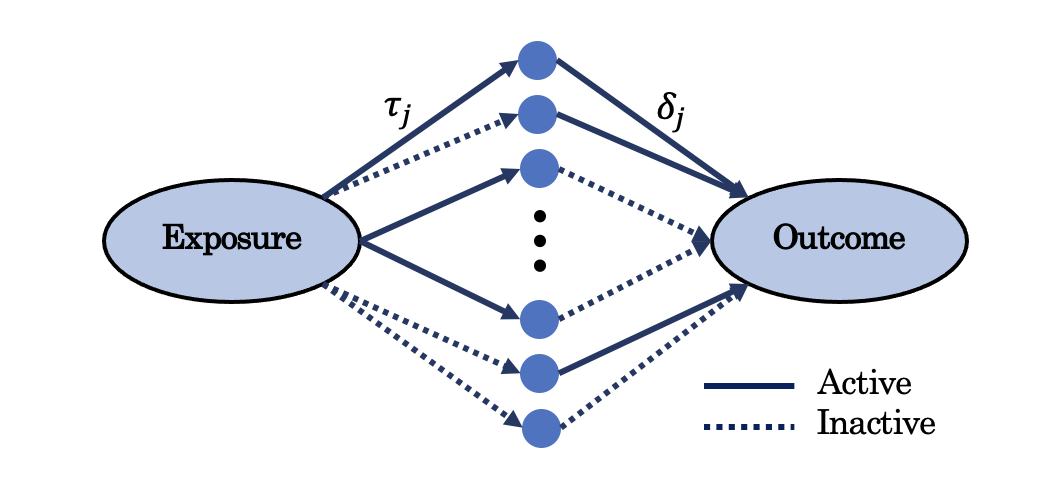}
\caption{A directed acyclic graph illustrating high-dimensional mediators. Each mediator has an independent pathway, with the treatment-mediator and mediator-outcome pathways able to independently activate or deactivate. A causal effect is transmitted through a mediator only when both pathways are active. A correlation structure among mediators is also permitted. 
} \label{fig:DAG}
\end{figure}

In the presence of high-dimensional mediators, we assume that each mediator operates through an independent pathway between the exposure and the outcome. As illustrated in Figure \ref{fig:DAG}, we adopt a sparsity assumption, where only a subset of the high-dimensional mediators have active pathways for both exposure-mediator and mediator-outcome relationships. Additionally, we allow for correlated (non-causal) relationships among the mediators.

In this setting, the target causal estimand is the active mediation effect, defined as the average effect from the mediation pathways where both pathways centered around each mediator are active. Formally, if the 
$j$-th mediator has an active pathway, the indirect effect for two different exposure levels $a$ and $a^\prime$ ($IE_j(a, a^\prime)$) through this pathway is defined as follows:
\begin{eqnarray}
IE_j(a,a^\prime) &=& E\left[Y_i(a, \{M_{1,i}(a), M_{2,i}(a),\ldots, ,M_{q,i}(a) \}) \right. \nonumber \\
 & & - \left. Y_i(a, \{M_{1,i}(a), M_{2,i}(a),\ldots, M_{j-1,i}(a), M_{j,i}(a^\prime),M_{j+1,i}(a), \cdots,M_{q,i}(a) \}) \right].  \nonumber 
\end{eqnarray}   
Furthermore, the sum of all indirect effects ($IE(a,a^\prime)$) through the active mediators and the direct effect ($DE(a,a^\prime)$) not mediated through any mediator are as follows:
\begin{eqnarray}
IE(a, a^\prime) &= &\sum_{j \in \mathcal{M}} IE_j(a, a^\prime), \nonumber \\
DE(a, a^\prime) &= &E[Y_i(a, \{M_{1,i}(a^\prime), M_{2,i}(a^\prime),\ldots, ,M_{q,i}(a^\prime) \})  \nonumber \\
    &&~~ -Y_i(a^\prime, \{M_{1,i}(a^\prime), M_{2,i}(a^\prime),\cdots, ,M_{q,i}(a^\prime) \})], \nonumber 
\end{eqnarray}
where $\mathcal{M}$ denotes the set of indices for mediators with active mediation pathways.

\subsection{Observed data model}\label{sec:ma1}
Let $\boldsymbol{X}_i$, $A_i$, $\boldsymbol{M}_i$, and $Y_i$ represent the $p$-dimensional vector of covariates, exposure, $q$-dimensional mediators, and the outcome for the $i$-th subject, respectively. The observed data models for the high-dimensional mediators and the outcome are specified as follows:
\beq
\boldsymbol{M}_i = \boldsymbol{\beta}_0 + \boldsymbol{\tau} A_i + B^\top \boldsymbol{X}_i + \epsilon_M, \quad \epsilon_M \sim \textrm{MVN}(\bzero, \Sigma), \label{eq:mediator} \\
Y_i =  \alpha_0 + \boldsymbol{\delta}^\top  \boldsymbol{M}_i + \boldsymbol{\alpha}^\top \boldsymbol{X}_i + \alpha_{p+1} A_i + \epsilon_Y, \quad \epsilon_Y \sim \textrm{Normal}(0, \sigma^2), \label{eq:outcome}
\eeq
where $\boldsymbol{\beta}_0 = (\beta_{0,1}, \ldots, \beta_{0,q})^\top$ denotes the intercepts, and $B$ is the $p \times q$ matrix with columns $\boldsymbol{\beta}_1, \ldots, \boldsymbol{\beta}_q$, where $ \boldsymbol{\beta}_j = (\beta_{j,1}, \ldots, \beta_{j,p})^\top$ for $j = 1, \ldots, q$. The vector $ \boldsymbol{\tau} = (\tau_1, \ldots, \tau_q)^\top$ captures the effects of exposure $A_i$ on the corresponding mediators, and $\Sigma$ is the variance-covariance matrix of a multivariate normal (MVN), allowing for associational relationships among the mediators. In the outcome model, $\boldsymbol{\delta} = (\delta_1, \ldots, \delta_q)^\top$ and $ \boldsymbol{\alpha} = (\alpha_1, \alpha_2, \ldots, \alpha_p)^\top$, with $\alpha_{p+1}$ capturing the direct effect of exposure $A_i$ on the outcome $Y_i$.

Key features of these observed data models are: 1) Each mediator is modeled marginally (i.e., without influencing each other), and in the outcome model, mediators are linearly connected without interactions, which aligns with the structure required to estimate the mediation effect through independent pathways, as assumed in Figure \ref{fig:DAG}; and 2) the use of a multivariate mediator model effectively captures the correlated nature of high-dimensional mediators.

\subsection{Identifying assumptions}\label{sec:ma2}

Linking potential outcomes to observed data requires additional assumptions beyond the SUTVA. While certain potential outcomes can be identified under SUTVA, outcomes involving multiple exposure levels represent unobservable quantities that cannot be identified from observed data without further assumptions.

To address the identification of such unobservable quantities, we used the assumption of causally independent mediators as proposed by \cite{imai2013identification}. In brief, this assumption states that each mediator is independent of the potential outcome, given the confounders. For further details, refer to \cite{imai2013identification}.This assumption enables the nonparametric identification of each $IE_j(a, a^\prime)$, as well as the overall indirect effect $IE(a, a^\prime)$ and the direct effect $DE(a, a^\prime)$ for any two different exposure levels $a$ and $a^\prime$. Building on the observed data model from Section 2.1, the mediation effect through the $j$-th mediator is redefined as $IE_j(a,a^\prime) = (a-a^\prime)\delta_j \tau_j$. Moreover, the sum of all indirect effects and the direct effect are represented as $IE(a, a^\prime) = (a-a^\prime)\sum_{j=1}^q \delta_j\tau_j$ and $DE(a, a^\prime) = \alpha_{p+1}(a-a^\prime)$.


\section{Bayesian Variable Selection}\label{sec:bvs}

In this section, we present prior distributions, including a novel multivariate prior for the variable selection indicators, specially designed to leverage the dependence structure among mediators in a high-dimensional mediation analysis framework. We also address potential practical challenges, such as specifying a key hyperparameter and selecting appropriate thresholding for variable selection, along with their corresponding solutions. Details of the computational algorithm are provided in Supplementary Materials A.

\subsection{Prior Distributions}\label{sec:bvs1}

The proposed Bayesian framework requires specification of prior distributions for the unknown model parameters. For automated variable selection, we employ spike-and-slab priors for the regression parameters $\btau$ in the mediator model and $\bdelta$ in the outcome model \citep{george1993variable, george1997approaches}. Specifically, following priors are considered for $\tau_j$ and $\delta_j$, $j=1, \cdots, q$: 
\beq
	\tau_j | \gamma_j, \Sigma &\sim& \gamma_j \textrm{Normal}(0, v_j^2 \Sigma_{(j,j)}) + (1-\gamma_j) I_0,  \nonumber \\
	\delta_j | \omega_j, \sigma^2 &\sim& \omega_j \textrm{Normal}(0, \psi_j^2 \sigma^2) + (1-\omega_j) I_0, \label{eq_ss}
\eeq
where $\gamma_{j}$ and $\omega_{j}$ are binary latent variables that determine the membership of $\tau_j$ and $\delta_j$ to one of the components in the mixture distribution in the priors, respectively. $v_j^2$ and $\psi_j^2$ are the hyperparameters to specify. The exposure is considered to have a causal pathway to the $j$-th mediator if the data support $\gamma_j=1$ over $\gamma_j=0$. A similar interpretation holds for $\delta_j$ in the outcome model. 

Independent and marginal Bernoulli priors are typically assumed for the latent binary variables $\gamma_j$ and $\omega_j$. However, use of the independent priors is less appealing in the mediation analysis in that, i) the information on the within-subject correlations among mediators in the data will not be utilized in variable selection in (\ref{eq:mediator}), despite the possibility that highly to moderately correlated mediators (see Figure \ref{fig:corr}) may be affected by the exposure; ii) variable selection will be performed separately in (\ref{eq:mediator}) and (\ref{eq:outcome}), leading to a subset of mediators with $\gamma_j=0$ being unnecessarily tested in (\ref{eq:outcome}). To address these limitations, we propose an innovative approach for specifying priors for $(\bgamma, \bomega)$ that integrate underlying dependence between mediators into the variable selection in (\ref{eq:mediator}), motivated by an MRF prior \citep{li2010bayesian}, and encourage the sequential selection of important intermediate variables from (\ref{eq:mediator}) for consideration in candidate mediators in (\ref{eq:outcome}), enabling more efficient and practical posterior simulations. Hereafter, we refer to these priors as MRF-SSB priors, with SSB representing sequential subsetting Bernoulli. Specifically, the conditional representation of the MRF-SSB priors is given by 
\beq
\pi(\gamma_j| \bgamma_{(-j)}, \Sigma) = p_j(\bgamma, \Sigma)^{\gamma_j}\left\{1-p_j(\bgamma, \Sigma)\right\}^{1-\gamma_j},  \nonumber \\
\pi(\omega_j) = \tilde{\gamma}_j\theta_{\omega}^{\omega_j}(1-\theta_{\omega})^{1-\omega_j} + (1-\tilde{\gamma}_j)I_0, \nonumber
\eeq
where $\bgamma_{(-j)}$ is the vector $\bgamma$ with the $j$-th element removed, $\tilde{\gamma}_j$ denotes the value of $\gamma_j$, and
\beq
p_j(\bgamma, \Sigma) = \left\{1 + \exp \left(-\theta_{\gamma} - \eta \sum_{r \ne j} |c_{r,j}|\gamma_r\right)\right\}^{-1}. \nonumber 
\eeq
$c_{r,j}$ is the correlation between the $r$-th and $j$-th mediators, calculated from $\Sigma$. It is noted that the SSB prior is conditional on the value $\tilde{\gamma}_j$, instead of the parameter $\gamma_j$, highlighting the ``sequential subsetting" procedure, which forms candidate mediators for consideration in the selection of mediator-outcome causal pathways, receives feedback solely from the exposure-mediator model, but not vice versa. This approach is similar to the cutting feedback method proposed by \cite{lunn2009combining}. While the posterior distribution is constructed through correct decomposition of model parameters, the posterior updating process  for $\bgamma$ does not incorporate the full conditional distribution.

The hyperparameter $\theta_{\gamma}$ controls the shrinkage of the model together with $v_j^2$ in (\ref{eq_ss}). The hyperparameter $\eta$ determines the extent to which other mediators, for a given mediator, impact the probability of including the exposure in the model (\ref{eq:mediator}). We note that setting $\eta$=$0$ leads to $\pi(\gamma_j | \bgamma_{(-j)}, \Sigma)$ being equivalent to the conventional independent Bernoulli (IB) prior for $\bgamma$. The hyperparameter $\theta_{\omega}$ represents the initial guess for the proportion of mediating variables in the outcome model. 
\begin{wrap}
{\red Specifically, the conditional representation of the proposed joint Markov random field (JMRF) prior is expressed as the product of two sequentially applied priors as follows:
\beq
\pi(\gamma_j, \omega_j | \bgamma_{(-j)}, \Sigma) &=& \pi(\gamma_j | \bgamma_{(-j)}, \Sigma)\pi(\omega_j | \gamma_j)  \nonumber \\
&=& p_j(\bgamma, \Sigma)^{\gamma_j}\left\{1-p_j(\bgamma, \Sigma)\right\}^{1-\gamma_j} \pi(\omega_j | \gamma_j) \nonumber
\eeq
where $\bgamma_{(-j)}$ is the vector $\bgamma$ with the $j$-th element removed and
\beq
p_j(\bgamma, \Sigma) &=& \left\{1 + \exp \left(-\theta_{\gamma} - \eta \sum_{r \ne j} |c_{r,j}|\gamma_r\right)\right\}^{-1}, \nonumber \\
\pi(\omega_j | \gamma_j) &=& \begin{cases} \theta_{\omega}^{\omega_j}(1-\theta_{\omega})^{1-\omega_j},& \gamma_j=1; \\[2\jot]
0, &\gamma_j=0. \end{cases} \nonumber
\eeq
}
\end{wrap}
We assign $\bbeta_0 \sim \textrm{MVN}(0, h_0 I_q)$, $\bbeta_j \sim \textrm{MVN}(0, c_0 I_p)$, $\alpha_0 \sim \textrm{Normal}(0, s_0)$, $\balpha \sim \textrm{MVN}(0, t_0 I_p)$, and $\alpha_{p+1} \sim \textrm{Normal}(0, k_0)$. Lastly, we use the conjugate hyperpriors inverse-Gamma($\nu_0/2$, $\nu_0 \sigma_0^2 / 2$) and inverse-Gamma($\nu_1/2$, $\nu_1 \sigma_1^2 / 2$), for $\sigma_\Sigma^2$ and $\sigma^2$, respectively. ($h_0$, $c_0$, $s_0$, $t_0$, $k_0$, $\nu_0$, $\sigma_0^2$, $\nu_1$, $\sigma_1^2$) are the hyperparameters to be specified.

\subsection{Practical considerations: phase transition and selection of variables}\label{sec:bvs2}

The Markov property of the proposed MRF-SSB priors is both conceptually and computationally advantageous, as it facilitates the characterization of the dependence structure among high-dimensional mediators. However, careful attention is required when setting values of ($\eta$, $\theta_{\gamma}$), since even small increases in these hyperparameters can significantly shift the size of the mediator model ($\sum_{j=1}^q\gamma_j$), a phenomenon known as \emph{phase transition} for MRF priors \citep{li2010bayesian}. 

To specify $\eta$ for a given value of $\theta_\gamma$, \cite{stingo2011incorporating} employ prior simulations over a grid of $\eta$ values to identify the phase transition boundary, denoted as $\eta_{pt}$. A Beta prior is then placed on the ratio $\eta/\eta_{pt}$. In this work, we adopt a modified version of the posterior simulation approach, as demonstrated by  \cite{lee2017multivariate}, which effectively addresses the relative underestimation of $\eta_{pt}$ in the prior simulation method. Specifically, we run the following algorithm to detect $\eta_{pt}$ over the posterior simulations: i) Simulate $M_{pt}$ posterior samples for model parameters over a grid of $\eta$ values, $\{\eta^g:g=0,\ldots,G, \eta^0=0\}$; ii) Let $\gamma_{(\eta^g)}$ denote the 50-th percentile of the posterior mean estimates of $\bgamma$ across mediators for the model (\ref{eq:mediator}) with $\eta=\eta^{g}$. Calculate $\gamma_{(\eta^g)}$ for $g=0,\ldots,G$; iii) Identify the phase transition value $\eta_{pt}$ as $\eta^g$ if $\gamma_{(\eta^g)}-\gamma_{(\eta^0)} > 0.05$; iv) Set $\eta=\eta^{g-1}$ in the proposed model.

Posterior inference for variable selection can be performed either through the joint posterior distribution of ($\bgamma$, $\bomega$), or via marginal posterior probabilities of inclusion (PPI) of the latent parameters. In our framework, We adopt the latter. Specifically, the indirect effect through mediator $j$ is considered significant if the corresponding PPI exceeds a threshold $\kappa$, i.e., $\textrm{PPI}_j=P(\gamma_j=1, \omega_j=1 | \mathcal{D}) > \kappa$. A commonly used cutoff value of $\kappa=0.5$ represents the median probability model. Given the sensitivity of this approach to the choice of ($\theta_\gamma$, $\theta_\omega$), we implement an adaptive thresholding method from \cite{wadsworth2017integrative}, where $\kappa$ is selected to control the Bayesian false discovery rate (FDR). Specifically, we find an optimal value of $\kappa$ such that FDR($\kappa$) is less than a pre-specified Bayesian FDR value (e.g. 0.05), where FDR($\kappa$)=$\sum_{j=1}^q(1-\textrm{PPI}_j) * I(\textrm{PPI}_j >\kappa)/\{\sum_{s=1}^q I(\textrm{PPI}_s >\kappa)\}$.

\subsection{Computational Scheme} \label{sec:computation}

We develop a computationally efficient algorithm based on the covariance structure of a factor analytic (FA) model \citep{hogan2004bayesian}, defined as $\Sigma=\sigma_\Sigma^2(\blambda\blambda^{\top}+I_q)$, where $\blambda=(\lambda_1,\ldots,\lambda_q)^{\top}$ is a $q\times 1$ vector of factor loadings. We assume a \textrm{Normal}($\mu_\lambda$, $h_\lambda \sigma_\Sigma^2 I_q$) prior with hyperparameters ($\mu_\lambda$, $h_\lambda$) for each $\lambda_j$'s. Although this FA covariance model is relatively restrictive, it ensures computational feasibility for high-dimensional data in the proposed mediation analysis framework. Moreover, when used for the multivariate variable selection method with the MRF prior, the FA model demonstrates considerable robustness even in cases when $\Sigma$ is misspecified \citep{lee2017multivariate}. In Section \ref{sec:simulation}, we examine the performance of our variable selection method under model misspecification scenarios.

Estimation and inference for the proposed framework are performed based on samples from the joint posterior distribution by either exploiting prior-posterior conjugacies or using the adaptive Metropolis-Hastings algorithm. To improve the convergence of the stochastic search variable selection (SSVS) steps for updating $\btau$ and $\bdelta$, we introduce ``refinement" steps, where the non-zero elements of the parameters are further updated via the Gibbs sampler. This refined SSVS algorithm shows substantially enhanced mixing of MCMC chains in our numerical studies, leading to improved computational performance. A detailed description of the proposed computational scheme is provided in Supplementary Materials A. To increase the computation speed, we develop a series of core functions in $\texttt{C++}$. The algorithm is publicly available on the first author's GitHub page (\href{https://github.com/YounghoBae/BVSMed}{https://github.com/YounghoBae/BVSMed}). It takes approximately 30 minutes to run 10,000 MCMC iterations on a 3.2 GHz Apple M1 MacBook Pro for the analysis of the metabolomics data from MBS/MLVS ($n$=466, $q$=298, $p$=3).

\section{Simulation Studies}\label{sec:simulation}

\begin{wrap}
{\blue use the following text to introduce the notations for the three models compared:
The fitted model is represented as a hyphenated combination of different underlying mediator models (MVN or independent Normal’s), along with various priors for selection indicators in the mediator model (MRF or independent Bernoulli’s (IBs)), as well as those in the outcome model (SSB). In our simulation studies, we considered three models: the proposed MVN-MRF-SSB and two competing models, MVN-IB-SSB and Normal-IB-SSB.
}
\end{wrap}

In this section, we present simulation studies to assess the performance of the proposed framework in variable selection and estimation across a range of scenarios. The primary goal is to compare the proposed MRF-SSB prior with its standard counterpart in mediation analysis involving high-dimensional mediators. Specifically, we investigate how performance varies with the magnitude of indirect effects, the correlation structure among mediators, and the impact of misspecification of covariance structures.

\subsection{Set-up and data generation }\label{sec:sim_setup}

We considered the four different scenarios. In Scenarios I-III, we generated samples of size $n=1,000$, with $q=300$ mediators and $p=5$ confounders under the proposed model outlined in Sections \ref{sec:mediationa analysis} and \ref{sec:bvs}. Each scenario assumes different level and structure of correlation among mediators by varying $\blambda$. Specifically, $\boldsymbol{\lambda}=0.35\cdot\bone_q$ induced a correlation of 0.1 among mediators in Scenario I and III, while independent mediators were assumed with $\boldsymbol{\lambda}=\bzero_q$ in Scenario II. In Scenario IV, we matched the data dimensions with those from the NHSII/HPFS ($n=466$, $q=298$, $p=3$), and set $\Sigma$ to the sample covariance matrix estimated from the data, representing a case of misspecified covariance structure.

In all four scenarios, confounders ($\boldsymbol{X}$) were independently generated from a $\text{Normal}(0, 1)$, while the exposure was assumed to follow a $\text{Normal}(\boldsymbol{l}^\top \boldsymbol{X}, 1)$, where $\boldsymbol{l} = (0.5, 0.2, 0.7, 0.4, 0.6)^\top$ for Scenarios I-III and $\boldsymbol{l} = (0.5, 0.2, 0.7)^\top$ for Scenario IV. This setting indicates that the outcome and exposure are confounded by $\boldsymbol{X}$.

The parameters regulating the effect sizes of the exposure-mediator pathways and the mediator-outcome pathways, were set as $\btau = (-0.12\cdot\bone_5, -0.08\cdot\bone_5, -0.04\cdot\bone_5, 0.04\cdot\bone_5, 0.08\cdot\bone_5, 0.12\cdot\bone_5, \bzero_{q-30})^{\top}$ and $\bdelta = (\bd \otimes \bone_6, \bzero_{q-30})^{\top}$, where $\bd=(0.5, 1.0, 1.5, 0, 0)^{\top}$ in Scenarios I, II, and IV. It is noted that, in Scenario IV, we rearranged the elements of $\Sigma$ so that highly correlated mediators (corr $>$ 0.6) are aligned with the first 30 exposure-mediator pathways in this data-generating mechanism. We considered the \emph{null} case by setting all elements of $\btau$ to zero, indicating no active pathways, in Scenario III. Finally, we set $\boldsymbol{\beta}_0=0.1\cdot\bone_q$, $\boldsymbol{B}=0.1\cdot\bone_{p\times q}$, and $\sigma^2_\Sigma=0.5$ in the mediator model (\ref{eq:mediator}), and  $\alpha_0=2$, $ \boldsymbol{\alpha}=2\cdot\bone_p$, $\sigma^2=0.5$, and $\alpha_{p+1}=2$ in the outcome model (\ref{eq:outcome}).

\subsection{Specification of hyperparameters and analysis settings}\label{sec_sim_setting}

We fitted three different models to 480 simulated datasets, 120 under each of the four scenarios. The models are the proposed MVN-MRF-SSB, and two competing models, Normal-IB-SSB and MVN-IB-SSB. These models represent different combinations of underlying mediator models (MVN or independent Normals) and priors for selection indicators in the mediator model (MRF or IB) and the outcome model (SSB). The simplest of the three, Normal-IB-SSB, assumes no correlation among mediators, similar to the approach by \cite{song2020bayesianOmics}. In contrast, MVN-IB-SSB explicitly models the correlation structure among the mediators, providing more flexibility, but it is limited in that it does not fully exploit the dependence of mediators in pathway identification, as it employs simple IB priors for selection indicator parameters. 

As outlined in Section \ref{sec:bvs}, the proposed Bayesian framework requires the specification of hyperparameters. The hyperparameters were set as $(h_0, c_0, s_0, t_0, k_0, h_{\lambda})=100$, $(\mu_0, \mu_1, \mu_{\lambda}) = 0$, $(\nu_0, \nu_1) = 6$, $(\sigma^2_0, \sigma^2_1) = 1/3$, to make the corresponding priors relatively non-informative. We set logit$^{-1}(\theta_\gamma)=\theta_\omega = 0.1$, indicating prior probabilities of 0.1 for both the association between the exposure and mediators in (\ref{eq:mediator}) and the association between mediators and the outcome in (\ref{eq:outcome}). Finally, $v_j^2=\psi_j^2=9$, and the value of $\eta$ for the proposed MVN-MRF-SSB was selected based on the posterior simulation approach introduced in Section \ref{sec:bvs2}. For Scenarios I-III, we ran each MCMC chain for 150,000 iterations, using the first half as burn-in. In Scenario IV, where the models may require additional iterations to adequately adapt to the complex underlying correlation structure and smaller sample size setting, we increased the total number of iterations to 300,000. We assessed model convergence via visual inspection of trace plots and further confirmed it by calculating the Gelman-Rubin potential scale reduction (PSR) statistic \citep{gelman2003bayesian} for representative model parameters.
\begin{wrap}
Convergence of the models was assessed by inspection of trace plots. 
\end{wrap}

We evaluated the performance of the variable selection feature of the models using five quantities: true positive rate (TPR), false positive rate (FPR), negative predictive value (NPV), positive predictive value (PPV), and the number of variables selected (NVS). The indirect effect estimates were summarized as the median of the posterior means from 120 replicates for each scenario, with the corresponding standard deviations also provided.

\begin{wrap}

\begin{itemize}
    \item common for scenario 1,2,3,4
    \begin{itemize}
        \item $\delta = (0.5, 1, 1.5, 0, 0)\times 6$ : 18 significant values
        \item $\theta_\gamma = -2.2, \theta_\omega = 0.1$
        \item $\mu_\lambda = 0, h_\lambda = 100$
        \item $\nu = 3, \psi = 3$
        \item $h_0 = c_0 = s_0 = t_0 = k_0 = 100$
        \item $V_\tau = V_\delta = V_\lambda = 0.01$
        \item $\mu_0 = \mu_1 = 0, \nu_0 = \nu_1 = 6, \sigma^2_0 = \sigma^2_1 = 1/3$
        \item 10 initial value for $\tau, \delta$ = 0 
    \end{itemize}
    \item scenario1
    \begin{itemize}
        \item iter/burn/thin = 150000/75000/75
        \item $\eta = 0$ for Normal-IB-SSB and MVN-IB-SSB. $\eta = 1.04$ for MVN-MRF-SSB
    \end{itemize}
    \item scenario2
    \begin{itemize}
        \item iter/burn/thin = 150000/75000/75
        \item $\eta = 0$ for Normal-IB-SSB and MVN-IB-SSB. $\eta = 0.1$ for MVN-MRF-SSB
    \end{itemize}
    \item scenario3
    \begin{itemize}
        \item iter/burn/thin = 150000/75000/75
        \item $\eta = 0$ for Normal-IB-SSB and MVN-IB-SSB. $\eta = 0.1$ for MVN-MRF-SSB
    \end{itemize}
    \item scenario4
    \begin{itemize}
        \item iter/burn/thin = 350000/175000/175
        \item $\eta = 0$ for Normal-IB-SSB and MVN-IB-SSB. $\eta = 0.3$ for MVN-MRF-SSB
    \end{itemize}
\end{itemize}
\end{wrap}

\begin{table}[htp]
    \caption{Four operating characteristics$^{\dag}$ (\%) and the number of variables selected (NVS) for the exposure-mediator pathways ($\gamma$) and the combined exposure-mediator-outcome pathways ($\gamma \times \omega$) in the multivariate normal models with independent Bernoulli and sequential subsetting Bernoulli priors (MVN-IB-SSB) and the proposed Markov random field and sequential subsetting priors (MVN-MRF-SSB) across four simulation scenarios described in Section \ref{sec:sim_setup}.}\label{tbl_sim_roc}
    \centering
    \scalebox{0.7}{
    \begin{tabular}{c c c r r c r r c r r c r r}
    \hline
    \multicolumn{2}{c}{} && \multicolumn{5}{c}{MVN-IB-SSB} && \multicolumn{5}{c}{MVN-MRF-SSB} \\
    \cline{4-8}  \cline{10-14}
    \textbf{Scenario} & \textbf{} && \multicolumn{2}{c}{$\gamma$} && \multicolumn{2}{c}{$\gamma \times \omega$} && \multicolumn{2}{c}{$\gamma$} && \multicolumn{2}{c}{$\gamma \times \omega$} \\
    \cline{4-5} \cline{7-8} \cline{10-11} \cline{13-14}
    \multicolumn{2}{c}{} && \textbf{Mean} & \textbf{(SD)} && \textbf{Mean} & \textbf{(SD)} && \textbf{Mean} & \textbf{(SD)} && \textbf{Mean} & \textbf{(SD)} \\
    \hline
                    & TPR &&  58.4 &  (5.3) &&  56.9 &  (7.3) &&  68.7 &  (5.9) && 68.1 & (7.7)  \\
                    & FPR &&   0.3 &  (0.3) &&   1.6 &  (0.5) &&   1.4 &  (0.8) &&  0.7 & (0.4)  \\
                  I & PPV &&  95.8 &  (4.1) &&  70.1 &  (8.2) &&  84.7 &  (7.0) && 86.9 & (8.1)  \\
                    & NPV &&  95.6 &  (0.5) &&  97.3 &  (0.5) &&  96.6 &  (0.6) && 98.0 & (0.5)  \\
                    & NVS &&  18.3 &  (1.7) &&  14.7 &  (1.6) &&  24.5 &  (2.9) && 14.1 & (1.2)  \\ \cline{1-14}
                    & TPR &&  59.8 &  (5.6) &&  58.9 &  (7.4) &&  59.8 &  (5.2) && 58.9 & (7.2)  \\
                    & FPR &&   0.2 &  (0.3) &&   0.0 &  (0.1) &&   0.3 &  (0.3) &&  0.0 & (0.1)  \\
                 II & PPV &&  96.6 &  (3.9) &&  98.9 &  (3.0) &&  96.3 &  (4.1) && 99.0 & (2.8)  \\
                    & NPV &&  95.7 &  (0.6) &&  97.4 &  (0.4) &&  95.7 &  (0.5) && 97.4 & (0.4)  \\
                    & NVS &&  18.6 &  (1.7) &&  10.7 &  (1.4) &&  18.6 &  (1.6) && 10.7 & (1.3)  \\ \cline{1-14}
                    & TPR &&    -- &     -- &&    -- &     -- &&    -- &     -- &&    -- &    -- \\
                    & FPR &&   0.1 &  (0.2) &&   0.1 &  (0.2) &&   0.1 &  (0.2) &&   0.1 & (0.2) \\
      III$^{\ddag}$ & PPV &&    -- &     -- &&    -- &     -- &&    -- &     -- &&    -- &    -- \\
                    & NPV && 100.0 &  (0.0) && 100.0 &  (0.0) && 100.0 &  (0.0) && 100.0 & (0.0) \\
                    & NVS &&   0.4 &  (0.7) &&   0.4 &  (0.7) &&   0.4 &  (0.7) &&   0.4 & (0.7) \\ \cline{1-14}
                    & TPR &&  24.1 & ( 9.0) &&  13.2 & ( 9.6) &&  45.2 & (12.7) &&  21.9 & (12.4) \\
                    & FPR &&   0.3 & ( 0.4) &&   0.3 & ( 0.3) &&   0.7 & ( 1.6) &&   0.2 & ( 0.2) \\
                 IV & PPV &&  92.5 & (10.8) &&  75.3 & (27.6) &&  91.2 & (11.8) &&  90.9 & (13.3) \\
                    & NPV &&  92.2 & ( 0.9) &&  94.7 & ( 0.6) &&  94.2 & ( 1.3) &&  95.2 & ( 0.7) \\
                    & NVS &&   7.9 & ( 3.1) &&   3.2 & ( 2.3) &&  15.4 & ( 6.8) &&   4.4 & ( 2.4) \\ \hline
    \multicolumn{14}{l}{\footnotesize NOTE: Throughout values are based on results from 120 simulated datasets.}\\    
    \multicolumn{14}{l}{\footnotesize$\dag$ true positive rate (TPR), false positive rate (FPR), negative predictive value (NPV), and} \\ 
    \multicolumn{14}{l}{\footnotesize ~~ positive predictive value (PPV)}\\         
    \multicolumn{14}{l}{\footnotesize$^{\ddag}$ TPR and PPV are not presented, as there are no active pathways.}       
    \end{tabular}}
    \label{tab:simulation}
\end{table}

\begin{table}[htp]
    \caption{Estimated indirect effects (IEs)$^\dag$ and marginal posterior probabilities of inclusion (PPI) for the multivariate normal models with independent Bernoulli and sequential subsetting Bernoulli priors (MVN-IB-SSB) and the proposed Markov random field and sequential subsetting priors (MVN-MRF-SSB) for Scenario I. Estimated direct effect (DE)$^\dag$ is also provided.}\label{tbl_sim_est}
    \centering
    \scalebox{0.7}{
    \begin{tabular}{r r r r c r r c r r c r r r r c r r c r r}
    \hline
    &\multicolumn{3}{c}{} && \multicolumn{2}{c}{MVN-IB-SSB} && \multicolumn{2}{c}{MVN-MRF-SSB} && &  \multicolumn{3}{c}{} && \multicolumn{2}{c}{MVN-IB-SSB} && \multicolumn{2}{c}{MVN-MRF-SSB} \\
     \cline{6-7}\cline{9-10}\cline{17-18}\cline{20-21} 
    &\multicolumn{3}{c}{True} && \multicolumn{2}{c}{IE$_j$${^\ddag}$} && \multicolumn{2}{c}{IE$_j$} && &  \multicolumn{3}{c}{True} && \multicolumn{2}{c}{IE$_j$} && \multicolumn{2}{c}{IE$_j$} \\
    $j$ & \textbf{$\tau_j$} & \textbf{$\delta_j$} & IE$_j$ && PM (SD) & $\overline{\text{PPI}}$ && PM (SD) & $\overline{\text{PPI}}$ && $j$ & $\tau_j$ & $\delta_j$ & IE$_j$ && PM (SD) & $\overline{\text{PPI}}$ && PM (SD) & $\overline{\text{PPI}}$ \\ \cline{1-10} \cline{12-21}
    1  & -0.12 & 0.5 & -0.12 && -0.19 (0.05) & 0.98 && -0.16 (0.04) & 0.99 && 16 & 0.04 & 0.5 &  0.04 &&  0.08 (0.04) & 0.14 &&  0.06 (0.04) & 0.31 \\
    2  & -0.12 & 1.0 & -0.24 && -0.32 (0.07) & 0.98 && -0.29 (0.06) & 0.99 && 17 & 0.04 & 1.0 &  0.08 &&  0.13 (0.05) & 0.08 &&  0.10 (0.06) & 0.26 \\
    3  & -0.12 & 1.5 & -0.36 && -0.44 (0.09) & 1.00 && -0.40 (0.08) & 1.00 && 18 & 0.04 & 1.5 &  0.12 &&  0.17 (0.08) & 0.07 &&  0.13 (0.08) & 0.18 \\
    4  & -0.12 & 0.0 &  0.00 && -0.06 (0.03) & 0.65 && -0.05 (0.02) & 0.14 && 19 & 0.04 & 0.0 &  0.00 &&  0.03 (0.02) & 0.05 &&  0.02 (0.01) & 0.05 \\
    5  & -0.12 & 0.0 &  0.00 && -0.06 (0.04) & 0.61 && -0.04 (0.03) & 0.15 && 20 & 0.04 & 0.0 &  0.00 &&  0.03 (0.02) & 0.10 &&  0.02 (0.01) & 0.06 \\
    6  & -0.08 & 0.5 & -0.08 && -0.13 (0.04) & 0.68 && -0.11 (0.04) & 0.80 && 21 & 0.08 & 0.5 &  0.08 &&  0.12 (0.04) & 0.54 &&  0.10 (0.03) & 0.81 \\
    7  & -0.08 & 1.0 & -0.16 && -0.20 (0.06) & 0.65 && -0.19 (0.05) & 0.82 && 22 & 0.08 & 1.0 &  0.16 &&  0.21 (0.07) & 0.70 &&  0.19 (0.06) & 0.85 \\
    8  & -0.08 & 1.5 & -0.24 && -0.29 (0.08) & 0.67 && -0.28 (0.08) & 0.81 && 23 & 0.08 & 1.5 &  0.24 &&  0.29 (0.09) & 0.62 &&  0.28 (0.08) & 0.83 \\
    9  & -0.08 & 0.0 &  0.00 && -0.04 (0.02) & 0.34 && -0.03 (0.02) & 0.11 && 24 & 0.08 & 0.0 &  0.00 &&  0.04 (0.02) & 0.37 &&  0.03 (0.02) & 0.18 \\
    10 & -0.08 & 0.0 &  0.00 && -0.05 (0.02) & 0.37 && -0.03 (0.02) & 0.18 && 25 & 0.08 & 0.0 &  0.00 &&  0.04 (0.02) & 0.38 &&  0.03 (0.02) & 0.21 \\
    11 & -0.04 & 0.5 & -0.04 && -0.07 (0.04) & 0.10 && -0.05 (0.03) & 0.20 && 26 & 0.12 & 0.5 &  0.12 &&  0.19 (0.05) & 0.98 &&  0.16 (0.04) & 0.99 \\
    12 & -0.04 & 1.0 & -0.08 && -0.12 (0.06) & 0.07 && -0.10 (0.05) & 0.23 && 27 & 0.12 & 1.0 &  0.24 &&  0.31 (0.07) & 0.98 &&  0.29 (0.06) & 1.00 \\
    13 & -0.04 & 1.5 & -0.12 && -0.17 (0.09) & 0.04 && -0.14 (0.08) & 0.19 && 28 & 0.12 & 1.5 &  0.36 &&  0.43 (0.09) & 0.97 &&  0.40 (0.08) & 0.99 \\
    14 & -0.04 & 0.0 &  0.00 && -0.03 (0.02) & 0.02 && -0.02 (0.01) & 0.03 && 29 & 0.12 & 0.0 &  0.00 &&  0.07 (0.03) & 0.66 &&  0.05 (0.02) & 0.15 \\
    15 & -0.04 & 0.0 &  0.00 && -0.03 (0.02) & 0.06 && -0.02 (0.01) & 0.05 && 30 & 0.12 & 0.0 &  0.00 &&  0.07 (0.03) & 0.59 &&  0.04 (0.03) & 0.16 \\
    \hline
    &  \multicolumn{3}{c}{True} && DE${*}$ & && DE & \multicolumn{11}{c}{\textbf{}}\\
    &  \multicolumn{3}{c}{DE}      && \multicolumn{2}{c}{PM (SD)}      && \multicolumn{2}{c}{PM (SD)}      & \multicolumn{11}{c}{\textbf{}}\\ 
    \cline{1-10}
    &  \multicolumn{3}{c}{4.00}    && \multicolumn{2}{c}{3.99 (0.25)} && \multicolumn{2}{c}{4.00 (0.19)}  & \multicolumn{11}{c}{} \\    
    \hline
    \multicolumn{21}{l}{\footnotesize NOTE: Throughout values are based on results from 120 simulated datasets.}\\
    \multicolumn{21}{l}{\footnotesize ~~~~~~~~~~ Results are shown for the first 30 mediators ($j = 1, \ldots, 30$) that are causally affected by the exposure.}\\      
    \multicolumn{21}{l}{\footnotesize$\dag$ IE: indirect effect, defined as $2 \tau \delta$, DE: direct effect, defined as $2\alpha_{p+1}$. The value ``2" represents the difference between two exposure levels, which}\\
    \multicolumn{21}{l}{\footnotesize ~~~ correspond to the 25th and 75th percentiles of the exposure distribution.}\\
   \multicolumn{21}{l}{\footnotesize$\ddag$ The medians of the posterior means (PM) and posterior standard deviation (SD) of IE$_{j}$ conditioning on $\gamma_{j}=1$ and $\omega_{j}=1$, the medians of the posterior} \\
   \multicolumn{21}{l}{\footnotesize ~~~ means of PPI ($\gamma_j\times\omega_j$) are computed.}   \\
    \multicolumn{21}{l}{\footnotesize$*$ The medians of the posterior means (PM) and posterior standard deviation (SD) of DE are computed.} 
    \end{tabular}}
\end{table}

\subsection{Results}\label{sec_sim_results}

As aforementioned, the primary aim of our numerical studies was to evaluate the improvement achieved by the proposed MRF-SSB priors compared to a conventional counterpart. To this end, we presented the comparative analysis of the MVN-MRF-SSB model alongside the MVN-IB-SSB model in the main manuscript, allowing for an assessment of both models under identical conditions, differing only in their prior structures for $\bgamma$. Results from Normal-IB-SSB are provided in Supplementary Materials B.2 for reference.

When the mediators were uncorrelated (Scenario II), both models exhibited nearly identical variable selection performance (Table \ref{tbl_sim_roc}). However, in scenarios with correlated mediators and active pathways (I, IV), the proposed MVN-MRF-SSB model outperformed the MVN-IB-SSB model in sensitivity. For example, in Scenario I, the MVN-MRF-SSB achieved a TPR of 68\%, compared to 57\% for the MVN-IB-SSB, in detecting exposure-mediator-outcome pathways ($\gamma\times\omega$). This difference is further illustrated in the estimated indirect effects (IEs) presented in Table \ref{tbl_sim_est}. Both models successfully identified the IEs involving the six largest exposure-mediator effects ($|\tau_j| = 0.12$). However, the MVN-MRF-SSB performed much better at detecting IEs linked to smaller-magnitude exposure-mediator associations, yielding PPIs greater than 0.80, in contrast to 0.70 or less for the competing model, for mediators 6-8 and 21-23, where $|\tau_j| = 0.08$. Thus, this improved performance was attributed to the MRF prior’s ability to identify exposure-mediator pathways (TPR of 69\% compared to 58\% with the IB priors, see Table \ref{tbl_sim_roc} for $\gamma$) by effectively exploiting mediator correlations.

When $n$ was smaller and the covariance model was misspecified (Scenario IV), the overall sensitivity for detecting causal pathways, which were held constant across scenarios, decreased as expected. However, the ability to detect exposure-mediator pathways improved more significantly (44\% versus 22\% for $\gamma$) due to the inclusion of stronger correlations among relevant mediators ($>$0.6) in the MRF prior. This improvement resulted in greater power to identify active mediating pathways (19\% versus 11\% for $\gamma\times\omega$). 

The proposed model achieved the same level of specificity or better when compared to the competing model. In the null case (Scenario III), both models successfully excluded inactive pathways, even with the presence of correlated mediators. In Scenarios I and IV, however, the proposed MRF prior identified more relevant candidate exposure-mediator pathways than the IB priors, although it slightly increased the FPR (1.4\% versus 0.3\% and 0.6\% versus 0.2\%, respectively; see Table 1). Then, the sparsity property of the proposed SSB priors encouraged the exclusion of irrelevant pathways in the outcome model, resulting in a lower FPR for the final quantities of interest, indirect effects, in the MVN-MRF-SSB model (0.7\% versus 1.6\% and 0.1\% versus 0.2\%, respectively; see Table 1).

In the estimation of active pathways (Table \ref{tbl_sim_est}), when the indirect effect size was large (e.g., IE$_j$ for $j = 1, 2, 3$), the proposed MVN-MRF-SSB produced effect estimates closest to the true values, while the competing model overestimated the effect magnitudes. Additionally, as the indirect effect size increased, the estimated PPI from the proposed model approached 1, indicating a high level of confidence in selecting active mediation pathways. In cases with null indirect effects (e.g., IE$_j$ for $j = 4, 5$), the competing model provided effect estimates further from zero than the proposed model, and the corresponding PPIs were estimated to be closer to 1.When the indirect effect sizes were similar (e.g., for the pairs $j = 2$ and $j = 8$ or $j = 1$ and $j = 13$), decreasing $\tau$ led to increased bias and uncertainty in effect estimation in the competing model, along with a lower average estimated PPI. However, the proposed model showed a smaller degree of bias and uncertainty compared to the competing model, particularly for moderate $\tau$ values (e.g., $\pm 0.08$), where it maintained relatively high estimated PPIs. This further demonstrates that the variable selection method based on the MRF prior performs effectively, even for smaller $\tau$ values, resulting in more accurate indirect effect estimations. Estimates of active pathways for Scenarios II-IV are provided in Supplementary Materials B.1.

In summary, across all scenarios, our simulation studies highlighted the importance of accounting for correlations among mediators. The proposed method effectively addressed this challenge by integrating the correlation structure through a multivariate model and by incorporating mediator interdependence into the stochastic search variable selection process using MRF-SSB priors.

\begin{wrap}
The results for the selected mediators are summarized in Table \ref{tab:simulation}, which presents outcomes for the exposure-mediator pathways ($\gamma$) and the combined results for the exposure-mediator-outcome pathways ($\gamma \times \omega$). In Scenario 1, where mediators have moderate correlation, the Normal-IB-SSB model, which uses an independent Normal assumption, demonstrates lower performance across all metrics. Adjusting for the correlation structure among mediators by using a Multivariate Normal (MVN-IB-SSB) leads to slight improvements across all metrics. In the scenario with 30 mediators containing signals, both models produce NVS values of 17-18 for $\gamma$ and TPRs slightly above 0.5. In contrast, the proposed method achieves a TPR exceeding 0.7 and an NVS of 34.83, aligning more closely with the true 30 exposure-mediator pathways with signals.

Table \ref{tbl_sim_est} presents the estimated direct effect (DE) and indirect effects (IE$_j$'s) for Scenario 1. The two exposure levels for effect estimation are set at $a = 1$ and $a^\prime = -1$, which approximately reflect the 25th and 75th percentiles of the exposure distribution. We report the median (PM) and standard deviation (SD) of the posterior means from the replicates, as well as the average posterior probability of inclusion ($\overline{\text{PPI}}$) for each exposure-mediator-outcome pathway across 120 replicated datasets, with an adaptive threshold applied. The true value for the direct effect $DE(a, a^\prime) = 2\alpha_{p+1} = 4.0$ is set, and while all three models yield similar posterior mean estimates, the proposed model demonstrates the lowest uncertainty, as evidenced by the smallest standard deviation of the posterior means.

For the indirect effects (IE), we present results for the first 30 mediators, where true signals are assigned along the exposure-mediator pathways. The true values for the indirect effects, $ IE_j(a, a^\prime) = (a - a^\prime) \tau_j \delta_j $, are generated to have values in the set $\{0.00, \pm 0.08, \pm 0.12, \pm 0.16, \pm 0.24, \pm 0.36\}$ based on combinations of 
different $\tau$ and $\delta$ values. 

\end{wrap}

\begin{wrap}
\subsection{Data generating process}\label{sec:sim1}
\begin{align}
    \boldsymbol{X} &\sim \textrm{MVN}(\boldsymbol{0}, \boldsymbol{I}_p) \\
    \boldsymbol{A} &\sim \textrm{Normal}(\boldsymbol{XL}, 1) \\
\end{align}

In this section, we evaluate the performance of our proposed model through simulation experiments.
Confounders were simply extracted from the standard multivariate normal distribution, and treatment was extracted from the normal distribution, which averages the linear combination of these confounders ,that expressed as $\boldsymbol{XL}$.
Both the intercept term of the mediators model and the beta's corresponding to the coefficients of the confounders were set to 0.1.
The variance part $\sigma^2_\Sigma$ of the Mediators model and the variance $\sigma^2$ of the outcome model are both set to 0.5.
Both the intercept term of the outcome model and the coefficients of the confounders are set to 2. However, the effect of treatment on y was set to 0.5.
The number of observations, mediators, confounders, and tau and lambda are set differently for each scenario, so we will mention them in the description of each scenario.

\subsection{Competing models}
As mentioned earlier, our proposed model considers correlations between mediators, and considers variable dependence based on this correlation when selecting variables to select significant mediators.
To show the outstanding performance of our model, we use the following two models as a competing model.

Model 1 is the most basic conceptual model, which does not reflect the correlation between mediators, and for this reason, variable dependence is not reflected in variable selection. In other words, the covariance matrix of the mediators model is considered as a diagonal matrix, and the prior for $\gamma$ also uses an independent Bernoulli prior, like the outcome model's $\omega$, instead of the MRF prior.

Like model1, model2 is a slightly more complex model that does not consider variable dependence in variable selection, but considers the correlation of overall mediators.

In summary, by comparing model1 and model2, we can understand how much performance it increases to first consider the correlation between mediators, and by comparing model2 with our proposed model, we can understand how considering variable dependence affects the final analysis results.

\subsection{Metrics}
\subsubsection{Operation characteristics}
To evaluate the performance of models in the simulation, we used several operation characteristics: True Positive Rate (TPR), False Positive Rate (FPR), Negative Predictive Value (NPV), Positive Predictive Value (PPV), and the number of variables selected. These metrics provide a comprehensive assessment of model effectiveness.
\begin{itemize}
    \item True Positive Rate (TPR, Sensitivity): TPR measures the proportion of actual positives that the model correctly identifies. A high TPR indicates that the model captures key variables effectively.
    \item False Positive Rate (FPR): FPR quantifies the proportion of actual negatives that are incorrectly predicted as positives. A low FPR shows the model avoids selecting irrelevant variables.
    \item Negative Predictive Value (NPV): NPV is the proportion of predicted negatives that are correctly identified. A higher NPV suggests reliable negative predictions.
    \item Positive Predictive Value (PPV): PPV measures the proportion of predicted positives that are true positives. A high PPV indicates confidence in the selected variables.
    \item Number of Variables Selected: The number of variables chosen is crucial, as selecting too many can lead to overfitting, while too few may overlook key predictors.
\end{itemize}

\subsubsection{Rationale and insights}
These metrics were chosen because they provide a detailed evaluation of model performance, especially in terms of variable selection. By using TPR, FPR, NPV, and PPV, we can assess how well each model balances the identification of important variables while avoiding irrelevant ones. A model with high TPR and PPV reliably identifies key variables, while a low FPR suggests effective exclusion of irrelevant variables. The number of variables selected helps determine whether the model strikes the right balance between complexity and parsimony. Together, these metrics offer a holistic view of each model’s ability to handle variable selection and improve the generalizability of the results.

\subsubsection{Estimation of effects}
For each scenario, 120 simulations were performed to estimate the effect. The final effect estimate was determined by taking the median of the 120 posterior means from these simulations. Alongside the median, the standard deviation (SD) of the 120 posterior means was also reported. This approach provides a robust measure of central tendency (median) while accounting for the variability in the posterior means (SD). The median is used to reduce the influence of outliers in the posterior distributions, while the SD offers insight into the dispersion and consistency of the estimates across the 120 simulations.

\subsection{Scenario settings}

In Scenario 1, 2, and 3, data were generated by fixing the number of observations($n=1000$), the number of mediators($q = 300$), and the number of confounders($p = 5$), respectively. Accordingly, $\boldsymbol{L} = (0.5, 0.2, 0.7, 0.4, 0.6)$. However, in the case of Scenario 4, since information was taken from the actual data and conducted, each number was matched with the actual data($n = 466, q = 298, p=3, \boldsymbol{L} = (0.5, 0.2, 0.7)$). 
In the case of tau values, significant signals were used for the first 30 mediators in scenarios 1, 2, and 3. In contrast, in Scenario 4, 30 locations were selected based on information obtained from real data, not the first 30, and a significant signal was given to this location. The effect size of the signal used in all scenarios used six values($\pm0.12, \pm0.08, \pm0.04$), and these values were set based on the effect size limit that can be found in Model 1, which is the simplest model, based on the performance of model 1 among the competing models.

Scenario 1 is the main setting for evaluating our model. In this setting, all values of lambda were fixed at 0.35, in other words, the correlations of all mediators were fixed at about 0.1 to generate data. 
In Scenario 2, the mediators were generated independently by setting the values of all lambda to 0, and through this, an analysis was conducted to find out whether our proposed model performs well even if there is no correlation between the mediators.
In the third scenario, all tau values are set to zero, eventually leading to no path from treatment to mediators in the mediators model, thereby evaluating the performance of our model in an environment where there is no indirect effect in the data.
In the last scenario, a sample variance matrix of mediators was obtained from real data and data was generated based on this. Based on the given sample correlation matrix, a mediators group with a correlation of 0.6 or more between mediators was extracted, and a signal was assigned to 30 of them.

In the results of the first scenario, model2 considering the correlation between mediators (TPR for combine = 0.569) performs slightly better than model1 (TPR for combine = 0.553) which does not. However, when using the MRF prior that considering variable dependence here, i.e., our proposed model improves the performance by a much greater margin (TPR for combine = 0.704). In particular, when NVS was also considered, TPR came out better despite selecting more variables in terms of gamma.

In the second scenario, in fact, model1 is the true model of this setting. The results confirm that all three models perform similarly, indicating that the proposed model produces good results even though there is no real correlation between the mediators.

In the third scenario, as in the second scenario, the three models show similar results. This also shows that the proposed model estimates well in null setting as well as the other two simpler models.

In the last scenario, the number of observations itself is smaller than other settings, and the correlation also imitates actual data, so overall, it shows a weak value compared to the results of other scenarios. However, even in this scenario, it can be seen that what we want to see, that is, the performance of the proposed model, is better than the other two models. In terms of combine, TPR, PPV, etc., show better values, and the actual number of selected variables came out close to the number of variables set by the proposed model with the most actual effect.

\end{wrap}

\section{Application to Metabolomics Data}\label{sec:application}

The Bayesian framework proposed in this research is motivated by the integration of epidemiological studies and blood biomarker analyses, aiming to investigate the relationship between long-term adherence to the Mediterranean diet and cardiometabolic health. Plasma metabolomes are explored as potential mediators of this relationship. The specific objectives of the research include the identification of metabolites that mediate the causal pathway between adherence to the Mediterranean diet and cardiometabolic outcomes and the estimation of the corresponding indirect effects.

\subsection{Health Professionals Follow-up Study and Nurses' Health Study II}\label{sec:app1}

We applied the proposed variable selection method to data from $n=466$ participants in the NHSII/HPFS substudies, including $q=298$ plasma metabolites. The exposure of interest ($A$) was Alternate Mediterranean Diet score, inverse-normal transformed (mean=-0.02, s.d.=0.98), which reflects adherence to the Mediterranean Diet \citep{shan2023healthy}. The outcome ($Y$), a cardiometabolic disease risk score (mean=9.0, s.d.=2.7), was calculated by summing the level of three biomarkers associated with the pathogenesis of cardiometabolic disease, with higher scores indicating a high risk. Metabolites, the potential mediators ($\bM$), with high skewness (absolute skewness $>$2) were log-transformed, and all metabolites were standardized to $z$-scores within each substudy \citep{kim2013statistical}. Missing metabolite data were imputed using the random forest imputation method \citep{wei2018missing}. Additional details of the data collection and processing are provided in Supplementary Materials C.

\begin{wrap}
{\blue 
The application analysis was based on two substudies within the Health Professionals Follow-up Study (HPFS) and Nurses' Health Study II (NHSII). The HPFS is an ongoing prospective cohort study of 51,529 US male health professionals initiated in 1986 \citep{rimm1991prospective}. The participants’ diet, lifestyle and health-related information was collected at baseline and updated biennially. The HPFS substudy consisted of 307 men aged 45–80 years free from coronary heart disease, stroke, cancer or major neurological disease \citep{wang2021gut}. The participants provided blood samples from 2011 to 2013. The NHSII is an ongoing prospective cohort study that enrolled 116,429 female registered nurses in 1989 \citep{bao2016origin}. The cohort collected participant information on diet, lifestyle and medication use, ascertained diseases at baseline, and updated the information biennially using mailed questionnaires. The NHSII substudy enrolled 233 participants free from coronary heart disease, stroke, cancer or major neurological disease \citep{huang2019mind}. All of the participants provided blood samples from 2013 to 2014. 

After excluding participants without dietary data, a total of 466 participants (263 from the HPFS and 203 from the NHSII) were included.

Dietary information was collected at baseline (1986  for the HPFS and 1991 for the NHSII) and updated every 4 years thereafter with validated semi-quantitative food frequency questionnaires (SFFQs) \citep{yuan2017validity, al2021reproducibility}. To represent long-term dietary intake, we calculated cumulative average of dietary intake for each participant by summing up the intake levels from all available FFQs from baseline to the blood collection date and then dividing the sum by the number of FFQs. Alternate Mediterranean Diet (AMED) score was derived to measure the degree of adherence to the Mediterranean Diet \citep{shan2023healthy}.

The plasma metabolomics profiling in these two studies was performed using high-throughput liquid chromatography-mass spectrometry (LC-MS) techniques at the Broad Institute of MIT and Harvard (Cambridge, MA). A detailed description of the metabolomics profiling method can be found elsewhere \citep{paynter2018metabolic, wang2024integration}. We excluded metabolites with an intraclass correlation coefficient $<0.3$ across blinded quality control replicates or with $\geq 25\%$ missing data, resulting in 298 metabolites for final analyses. 

We derived a composite score to summarize the levels of biomarkers representing three well-established mechanisms underlying the pathogenesis of cardiometabolic diseases: LDL-C (‘bad’ cholesterol) for dyslipidemia, the ratio of TAG to HDL-C for insulin resistance, and hs-CRP for inflammation. Plasma concentrations of these biomarkers were measured using standard methods. To calculate the score, participants were first categorized into quintiles based on each biomarker level, ranked from lowest to highest, and assigned scores from 1 to 5. The cardiometabolic disease risk score was then calculated by summing these components, with higher scores reflecting a higher risk of cardiometabolic disease. 
\end{wrap}

\subsection{Specification of hyperparameters and analysis settings}

We fitted the proposed MVN-MRF-SSB and two competing models, MVN-IB-SSB and Normal-IB-SSB, to the HPFS/NHSII data. In our analyses, we included age (mean=66.7, s.d.=6.5), sex (0$=$male: 56.4\%; 1$=$female: 43.6\%), and body mass index (BMI, mean=26.0, s.d.=4.8 kg/m$^2$) as covariates ($\bX$, $p=3$) without performing variable selection on them. Hyperparameters for all models were set to the same values as in Section \ref{sec_sim_setting}. Three independent MCMC chains were run for a total of 300,000 iterations each, with the first half discarded as burn-in. Convergence was assessed by inspection of trace plots as well as calculation of the PSR statistic, requiring PSR values below 1.05 for model parameters.

\begin{table}[htp]
    \centering
    \caption{Estimated indirect effects (IEs)$^\dag$ of adherence to the Mediterranean diet on cardiometabolic disease risk through active pathways identified by multivariate normal models with independent Bernoulli and sequential subsetting Bernoulli priors (MVN-IB-SSB) and the proposed Markov random field and sequential subsetting priors (MVN-MRF-SSB). Estimated effects of diet on metabolites ($\tau$) are also provided.  }\label{tab_data_est}
    \scalebox{0.8}{
    \begin{tabular}{c c c c c c c c c}
    \hline
    && \multicolumn{3}{c}{MVN-IB-SSB} && \multicolumn{3}{c}{MVN-MRF-SSB} \\
    \cline{3-5} \cline{7-9}
    && \textbf{$\tau$} && \textbf{IE} && \textbf{$\tau$} && \textbf{IE} \\
    \cline{3-3} \cline{5-5} \cline{7-7} \cline{9-9}
    \textbf{Metabolites}$^\ddag$ && \textbf{PM (95\% HPDI)}$^*$ && \textbf{PM (95\% HPDI)}$^{**}$ && \textbf{PM (95\% HPDI)}$^*$ && \textbf{PM (95\% HPDI)}$^{**}$ \\
    \hline
    C53:2 TG  && -                    && -                   && -0.22 (-0.26,  0.00) && -0.22  (-0.35, 0.00) \\
    C51:3 TG  && -                    && -                   && -0.25 (-0.30, -0.21) &&  0.15  ( 0.00, 0.25) \\
    C51:2 TG  && -0.25 (-0.30,  0.00) &&  0.20 (-0.15, 0.47) && -0.25 (-0.30, -0.20) &&  0.42  ( 0.00, 0.71) \\
    C51:1 TG  && -0.25 (-0.28,  0.00) && -0.21 (-0.31, 0.00) && -0.24 (-0.30, -0.19) && -0.24  (-0.38, 0.00) \\
    C50:4 TG  && -                    && -                   && -0.21 (-0.24,  0.00) && -0.22  (-0.38, 0.00) \\
    C50:3 TG  && -0.23 (-0.27,  0.00) && -0.39 (-0.51, 0.00) && -0.24 (-0.28, -0.18) && -0.35  (-0.51, 0.00) \\
    C49:3 TG  && -0.29 (-0.34, -0.24) &&  0.11 (-0.23, 0.43) && -0.29 (-0.33, -0.24) &&  0.30  ( 0.00, 0.48) \\
    C38:4 GPC && -0.27 (-0.31, -0.22) &&  0.11 (-0.09, 0.22) && -0.27 (-0.32, -0.23) &&  0.13  ( 0.00, 0.22) \\
    C34:0 PS  &&  0.14 ( 0.00,  0.18) &&  0.09 ( 0.00, 0.15) &&  0.14 ( 0.00,  0.18) &&  0.08  ( 0.00, 0.12) \\
    C22:6 LPC && -                    && -                   &&  0.18 ( 0.13,  0.22) && -0.08  (-0.13, 0.00) \\
    C22:6 CE  && -                    && -                   &&  0.22 ( 0.17,  0.27) &&  0.16  ( 0.00, 0.26) \\
    C20:5 LPC &&  0.17 ( 0.12,  0.21) && -0.12 (-0.18, 0.00) &&  0.17 ( 0.12,  0.21) && -0.11  (-0.16, 0.00) \\
    C20:5 CE  && -                    && -                   &&  0.20 ( 0.16,  0.24) &&  0.10  ( 0.00, 0.16) \\
    C18:1 SM  && -                    && -                   && -0.16 (-0.20,  0.00) && -0.06  (-0.08, 0.00) \\
    C18:0 CE  && -0.16 (-0.20,  0.00) && -0.19 (-0.24, 0.00) && -0.16 (-0.20,  0.00) && -0.18  (-0.23, 0.00) \\
    \hline
    \multicolumn{9}{l}{\footnotesize NOTE: We adjusted for participants’ age, sex, and body mass index, potential confounders, but did not perform variable selection on them.}\\
    \multicolumn{9}{l}{\footnotesize$\dag$ IE: indirect effect, defined as $1.03 \tau \delta$. The value ``1.03" represents the difference between two exposure levels, which correspond to }\\
    \multicolumn{9}{l}{\footnotesize ~~~ the 30th and 70th percentiles of the exposure distribution.}\\        
    \multicolumn{9}{l}{\footnotesize $\ddag$ CE: cholesterol, TG: triacylglycerol, LPC: lysophosphatidylcholine, GPC: glycerophospholipid, SM: sphingomyelin,}\\
    \multicolumn{9}{l}{\footnotesize ~~~ PS: phosphatidylserine}\\   
    \multicolumn{9}{l}{\footnotesize $*$ Posterior median (PM) of $\tau$ (conditioning on $\gamma=1$) and 95\% highest posterior density intervals (HPDI) for $\tau$}\\
    \multicolumn{9}{l}{\footnotesize $**$ Posterior median (PM) of $1.03\tau\delta$ (conditioning on $\gamma=1, \omega=1$) and 95\% highest posterior density intervals (HPDI) for $1.03\tau\delta$}  
    \end{tabular}}
\end{table}

\begin{figure}[htp]
\centering
\includegraphics[width=14cm]{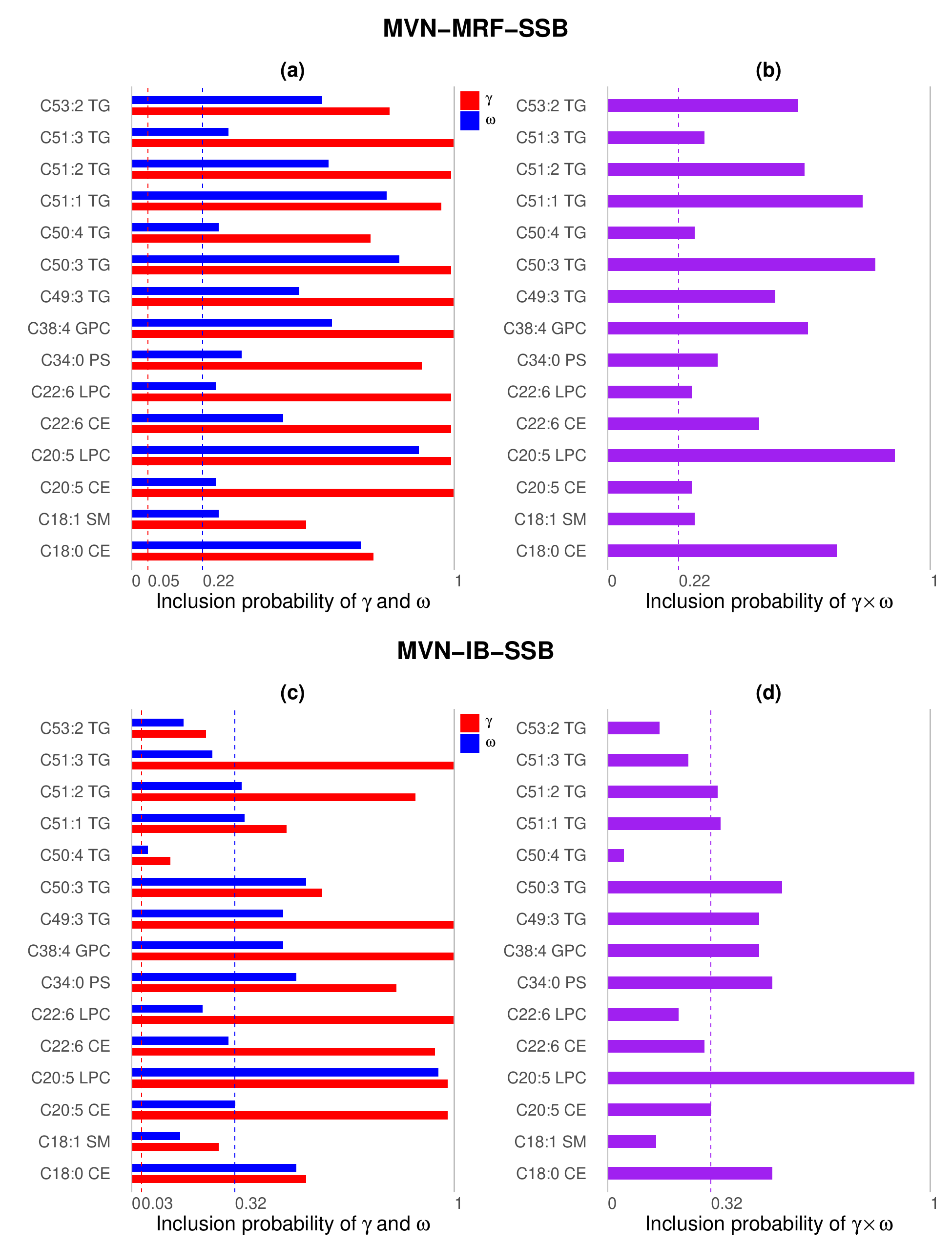}
\caption{Estimated marginal posterior probabilities of inclusion (PPI) associated with the indirect effects of adherence to the Mediterranean diet on cardiometabolic disease risk, estimated by multivariate normal models with independent Bernoulli and sequential subsetting Bernoulli priors (MVN-IB-SSB) and the proposed Markov random field and sequential subsetting priors (MVN-MRF-SSB). The vertical line represents the cutoff value $\kappa$ for variable selection, as determined by the adaptive thresholding method described in Section \ref{sec:bvs2}.} \label{fig_data_ppi}
\end{figure}

\begin{wrap}
\begin{itemize}
    \item $n = 466, q = 298, p = 3$
    \item iteration = 300000, burn.in = 150000, thin = 150
    \item $\theta_\gamma = -2.2, \theta_\omega = 0.1$
    \item $\mu_\lambda = 0, h_\lambda = 100$
    \item $\nu = 3, \psi = 3$
    \item $h_0 = c_0 = s_0 = t_0 = k_0 = 100$
    \item $V_\tau = V_\delta = V_\lambda = 0.01$
    \item $\mu_0 = \mu_1 = 0, \nu_0 = \nu_1 = 6, \sigma^2_0 = \sigma^2_1 = 1/3$
    \item $\eta = 0$ for Normal-IB-SSB and MVN-IB-SSB. $\eta = 0.2$ for MVN-MRF-SSB
    \item 10 initial value for $\tau, \delta$ = (-0.6, -0.4, -0.2, 0, 0.2, 0.4, 0.6, 0.8, 1, 1.2)
    \item select 3 chains between 10 chains that the combiantion with the smallest Gelman-Rubin statistics
\end{itemize}
\end{wrap}

\subsection{Results}\label{sec:app2}

As in the simulation studies, the main paper presented results from the proposed MVN-MRF-SSB model and the competing MVN-IB-SSB model. Results from the Normal-IB-SSB model were provided in Supplementary Materials D. We also provided overlaid trace plots for the illustrative parameters ($\blambda$, $\bdelta$) in the proposed model, which exhibited the slowest convergence among all model parameters, in Table D.2 and D.3. The plots indicated that the chains mixed well and converged early in the iterations, suggesting that, in practice, 300,000 iterations were not necessary for fitting our proposed model.

We applied a PPI cutoff of $\kappa$ for variable selection, estimated using the adaptive thresholding approach described in Section \ref{sec:bvs2}. The estimated inclusion probabilities for selected pathways, along with their corresponding metabolite mediators, were presented in Figure \ref{fig_data_ppi}. Among the 298 potential pathways, the competing method identified a total of 8 pathways. In contrast, the proposed MVN-MRF-SSB selected 7 additional pathways, highlighting 15 metabolites that may play an important intermediate role in the causal pathway between the Mediterranean diet and cardiometabolic disease risk. These results were attributed to the MRF-SSB priors devised for the proposed framework. As stated in Sections \ref{sec:bvs1} and \ref{sec_sim_results}, the MRF prior first improved the ability to identify subtle effects of exposure on a metabolite measure when other correlated metabolites were also affected by the exposure. This was evident when comparing the PPI of $\gamma$ for the two models, as shown in Figure \ref{fig_data_ppi}(a) and \ref{fig_data_ppi}(c). Subsequently, the SSB prior finalized the selection of the active pathways (Figure \ref{fig_data_ppi}(b) and \ref{fig_data_ppi}(d)). 

We provided estimated indirect effects and associated uncertainties as posterior medians (PM) and 95\% highest posterior density intervals (HPDI) in Table \ref{tab_data_est}. The two exposure levels ($a$ and $a^\prime$) for analyzing the effect were set at -0.54 and 0.49, corresponding to the 30th and 70th percentiles of the exposure, Alternative Mediterranean Diet score. Therefore, the $j$-th indirect effect $IE_j$ to be estimated is $1.03\tau_j \delta_j$, and the direct effect is $1.03\alpha_{p+1}$. The indirect effect estimates for identified pathways ranged from -0.35 to 0.42 in the MVN-MRF-SSB model, compared to a range of -0.39 to 0.21 in the MVN-IB-SSB model. Among the metabolites selected by both models, C51:2 triacylglycerols (TG) and C49:3 TG exhibited stronger indirect effects in the MVN-MRF-SSB model (PM: 0.42 and 0.30, respectively) than in the MVN-IB-SSB model (PM: 0.20 and 0.11, respectively). The estimated uncertainties, as indicated by the width of the HPDI, were largely comparable across the pathways identified by both models. The estimated direct effect was similar between the two models: PM (95\% HPDI) was -0.12 (-0.34, 0.11) for the MVN-MRF-SSB model and -0.14 (-0.35, 0.11) for the MVN-IB-SSB model. 

The results were consistent with previous studies that have shown protective indirect effects of the Mediterranean diet on cardiometabolic disease risk through several specific metabolites. For example, both models identified C20:5 lysophosphatidylcholine (LPC) (Table \ref{tab_data_est}), a well-established metabolic marker of fish intake, a key element of the Mediterranean diet. However, another marker associated with fish intake---C22:6 LPC---was uniquely identified by the proposed MVN-MRF-SSB model. These metabolites, which were inversely associated with cardiometabolic disease risk, further supported the protective (negative) indirect effects of the Mediterranean diet on cardiometabolic health. 


\begin{wrap}
{\red 
Table \ref{tab:results} presents the results for the indirect effect estimates. Results are shown only for the mediators (HMDB??) selected as having an active pathway in the proposed method. The IB prior method identifies no pathways beyond those found by the MRF prior method. From the posterior probability of inclusion plot in Figure \ref{fig:PIPmain}, the MRF prior is observed to select more exposure-mediator pathways. A total of 15 mediators are selected through the variable selection approach, with indirect effect estimates for these pathways ranging from -0.328 to 0.403. Notably, XX, YY, and ZZ (based on their 95\% credible intervals) show significant indirect effects, aligning with the roles of plasma metabolomes discussed in previous literature. The direct effect is estimated as XX (with a 95\% credible interval), which could indicate a direct effect of the Mediterranean diet on cardiometabolic health without involving metabolomes. However, it may also represent a pathway effect involving mechanisms not addressed in this analysis. In other words, it could reflect an effect mediated by other factors not considered here (do we have any?).
THIS NEEDS TO BE REFINED ONCE WE HAVE MORE INFORMATION ON THE RESULTS AND THE CONTEXT AS WELL.
}

\begin{table}[htp]
    \centering
    \caption{Estimated indirect effects of adherence to the Mediterranean diet on cardiometabolic disease risk through active pathways identified by multivariate normal models with independent Bernoulli and sequential subsetting Bernoulli priors (MVN-IB-SSB) and the proposed Markov random field and sequential subsetting priors (MVN-MRF-SSB). We adjusted for participants’ age, sex, and body mass index, potential confounders, but did not perform variable selection on them.}\label{tab_data_est}
    \scalebox{0.8}{
    \begin{tabular}{c c c c c c}
    \hline
    \textbf{} & \textbf{} && \textbf{MVN-IB-SSB} && \textbf{MVN-MRF-SSB} \\
    \cline{4-4} \cline{6-6}
    \textbf{HMDB} & \textbf{Metabolites}$^\dag$ && \textbf{PM(95\% CI)} && \textbf{PM(95\% CI)} \\
    \hline
    HMDB0042196 &              C53:2 TG && -                    && -0.22 (-0.37, -0.05) \\
    HMDB0042104 &              C51:1 TG && -0.21 (-0.37, -0.06) && -0.24 (-0.41, -0.08) \\
    HMDB0042103 &              C49:3 TG &&  0.11 (-0.31,  0.46) &&  0.27 (-0.02,  0.58) \\
    HMDB0012356 &              C34:0 PS &&  0.08 ( 0.02,  0.18) &&  0.08 ( 0.01,  0.16) \\
    HMDB0012101 &              C18:1 SM && -                    && -0.06 (-0.11, -0.02) \\
    HMDB0011701 &              C51:3 TG && -                    &&  0.15 ( 0.00,  0.31) \\
    HMDB0011252 &             C38:4 GPC &&  0.12 (-0.09,  0.23) &&  0.13 (-0.01,  0.26) \\
    HMDB0010404 &             C22:6 LPC && -                    && -0.07 (-0.18,  0.10) \\
    HMDB0010397 &             C20:5 LPC && -0.11 (-0.18, -0.05) && -0.11 (-0.17, -0.04) \\
    HMDB0010368 &              C18:0 CE && -0.18 (-0.26, -0.11) && -0.17 (-0.25, -0.09) \\
    HMDB0008731 &              C40:9 PC && -                    && -0.13 (-0.31,  0.04) \\
    HMDB0006733 &              C22:6 CE &&  0.08 (-0.17,  0.25) &&  0.17 ( 0.04,  0.34) \\
    HMDB0005435 &              C50:4 TG && -                    && -0.25 (-0.52, -0.01) \\
    HMDB0005433 &              C50:3 TG && -0.38 (-0.56, -0.20) && -0.33 (-0.58, -0.07) \\
    HMDB0005362 &              C51:2 TG &&  0.19 (-0.18,  0.64) &&  0.40 ( 0.03,  0.78) \\
    \hline
    \multicolumn{6}{l}{\footnotesize $\dag$CE: cholesterol, TG: triacylglycerol, PC: phosphatidylcholine, LPC: lysophosphatidylcholine,}\\
    \multicolumn{6}{l}{\footnotesize ~~GPC: glycerophospholipid, SM: sphingomyelin, PS: phosphatidylserine}
    \end{tabular}}
\end{table}

\end{wrap}

\section{Discussion}\label{sec:discussion}

In this paper, we introduced a novel Bayesian variable selection method for mediation analysis that accommodates the correlation structure of high-dimensional mediators through the priors of the selection indicators. We developed the special MRF-SSB priors to improve the power to detect active exposure-mediator-outcome pathways. The proposed mediation analysis approach allowed us to identify a subset of metabolites that mediate the relationship between adherence to a Mediterranean diet and cardiometabolic health outcomes while also enabling us to estimate and make inferences about the indirect effects associated with these metabolites. Notably, the proposed Bayesian framework identified a set of additional metabolites that were not detected by conventional Bayesian models using standard priors.

It is noted that, in practice, the estimated direct effects in our application may partially reflect effects mediated by other factors, such as filtered or unmeasured metabolites. Additionally, our data analyses were intended primarily as illustrative examples to highlight the potential advantages of the proposed method. Questions may arise regarding the analysis setup, as adjustments were made for only three confounders: age, sex, and BMI. We are currently investigating the performance of the methods across various model setups, including those with more comprehensive confounder adjustments.

Central to the proposed method is the incorporation of mediator correlations into the variable selection and the estimation of active pathways, which was shown to be effective in our numerical studies. However, it is important to acknowledge two limitations. First, while the proposed method exploits the correlation structure among high-dimensional mediators, it does not account for the sequential order among them (i.e., cause-and-effect relationships). When such a sequential order exists, defining individual indirect effects and introducing the corresponding identifying assumptions becomes a challenging problem, even with a limited number of mediators \citep{vanderweele2014mediation,daniel2015causal}. Extending this to high-dimensional mediators presents an additional topic for further research. One possible approach is to begin by estimating the directed acyclic graph (DAG) to identify active pathways \citep{castelletti2020bayesian}, then calculate each mediation effect as the indirect effect remaining after accounting for earlier mediators in the sequential pathway order. However, DAG estimation is particularly challenging with high-dimensional mediators, and developing precise identifying assumptions for the inferred DAG adds further complexity to the process. 

Second, the proposed model is based on the covariance structure of a FA model, which may be too restrictive for accurately capturing the complex dependencies among mediators in practice. Our simulation studies demonstrated improved performance of the proposed model, even when this structure was misspecified. Nevertheless, adopting more flexible covariance structures, such as the unstructured model or sparse infinite factor models \citep{bhattacharya2011sparse}, could further enhance the model's performance by leveraging the synergy between better-estimated correlations and the MRF-SSB priors. These extensions, however, require addressing computational challenges that must arise from estimating higher-dimensional models, such as consideration of regularization methods for sparse unstructured covariance models \citep{chu2013bayesian}. The development of more computationally efficient algorithm for detecting phase transitions is also necessary to scale up the method \citep{zhao2024multivariate}.

The proposed MRF-SSB priors can be adopted to mediation models with various types of mediators and outcomes. For non-continuous mediators, such as binary, categorical, or count data, generalized linear models can be incorporated into the mediator model. We are currently extending the framework to handle discrete and time-to-event outcomes, which are crucial for analyzing commonly observed outcomes in integrated epidemiological studies, such as disease status and time to diagnosis or death. These extensions are expected to improve the framework’s applicability in medical and epidemiological research by accommodating more complex data types.

In conclusion, the proposed Bayesian framework provides researchers with more powerful statistical tools to identify active pathways and estimate their associated effects in mediation analysis. The methods developed in this research along with publicly available software and insights from our numerical studies will benefit a wide range of fields beyond metabolomics study, particularly in contexts involving correlated continuous mediators.



\section*{Funding}

This work was supported by the National Institute of Dental and Craniofacial Research of USA (R03DE027486), the National Institute of General Medical Sciences of USA (R01GM126257) and the National Research Foundation (NRF) grant of South Korea (NRF-2020R1F1A1A01048168, NRF-2022R1F1A1062904, RS-2024-00407300). The Health Professionals Follow-Up study and Nurse’s Health Study were supported by the National Cancer Institute of USA (U01CA167552, U01CA176726).



\bibliographystyle{apalike}
\bibliography{BVSmedation}

\end{document}